\documentclass[a, full]{iucr}
\usepackage{amsmath}
\usepackage{amsthm}
\usepackage{multirow}
\usepackage{mathrsfs}
\usepackage{xcolor}
\usepackage{bm}
\usepackage{booktabs}
\usepackage[normalem]{ulem}
\usepackage{hyperref}
\usepackage{lineno}
\usepackage{soul}

\journalcode{J}

\begin{document}


\title{Quantifying Resolution in Pink Beam Dark Field X-ray Microscopy: Experiments and Simulations }

\author[a]{M.}{La Bella}
\author[a]{H. F.}{Poulsen}
\author[b,a]{S.}{Staeck}
\author[a]{N.A.}{Henningsson}
\author[c,d]{M.P.}{Kabukcuoglu}
\author[b]{C.}{Detlefs}
\cauthor[b]{C.}{Yildirim}{can.yildirim@esrf.fr}{}

\aff[a]{Technical University of Denmark, 2800 \city{Kgs Lyngby}, \country{Denmark}} 
\aff[b]{European Synchrotron Radiation Facility, 38043 \city{Grenoble}, \country{France}}
\aff[c]{Institute for Photon Science and Synchrotron Radiation (IPS), Karlsruhe Institute of Technology (KIT), 76344 \city{Eggenstein-Leopoldshafen}, \country{Germany}}
\aff[d]{Leibniz-Institut für Kristallzüchtung (IKZ), 12489 \city{Berlin}, \country{Germany}}

\date{\today}

\maketitle


\begin{abstract}


Pink-beam Dark-Field X-ray Microscopy (pDFXM) is a powerful emerging technique for time-resolved studies of microstructure and strain evolution in bulk crystalline materials. In this work, we systematically assess the performance of pDFXM relative to monochromatic DFXM when using a compound refractive lens (CRL) as the objective. Analytical expressions for the spatial and angular resolution are derived and compared with numerical simulations based on geometrical optics and experimental data. The pink-beam configuration provides an increased diffraction intensity depending on the deformation state of the sample, accompanied by a general tenfold degradation in angular resolution along the rocking and longitudinal directions. This trade-off is disadvantageous for axial strain mapping, but can be advantageous in cases where integrated intensities are needed. For a perfect crystal under parallel illumination with a pink beam, our results show that chromatic aberration is absent, whereas under condensed illumination it becomes significant. The aberration is shown to depend strongly on the local distortion of the crystal. Weak-beam imaging conditions, such as those required for resolving dislocations, are shown to remain feasible under pink-beam operation and may even provide an improved signal-to-noise ratio. The higher incident flux, enhanced by nearly two orders of magnitude, is quantified in terms of beam heating effects, and implications for optimized scanning protocols are discussed.

\end{abstract}


\section{Introduction}

Dark-Field X-ray Microscopy (DFXM) has emerged as a powerful full-field imaging technique for the non-destructive three-dimensional (3D) mapping of strain, orientation, and defect structures within bulk crystalline materials \cite{Simons2015,Poulsen2017}. By selectively imaging Bragg-diffracted X-rays, DFXM enables high-resolution characterization of microstructural features such as dislocations, domain structures, and strain gradients with sub-micrometer spatial resolution and milliradian angular sensitivity. The technique has been instrumental in studying a wide range of materials science phenomena, including plastic deformation, nucleation and growth studies, structural and magnetic phase transformations, ferroelectric behavior, biominerals and imaging of acoustic waves \cite{Ahl2017, Simons2018, Cook2018, Mavrikakis2019, Bucsek2019, Dresselhaus2021, Yildirim2023, Holstad2023, Lee2025, Gursoy2025, Zelenika2025, Zhou2025}. DFXM is often applied in combination with other diffraction imaging techniques, allowing multiscale characterization of hierarchically organized materials \cite{Gustafson2020, Gustafson2023, Chen2023, Lee2024MultiscaleDiffraction, shukla2025}.

The first implementations of DFXM have relied on monochromatic X-ray beams, providing well-defined reciprocal space selectivity and high angular resolution. This makes monochromatic DFXM particularly suitable for resolving subtle orientation variations in bulk crystalline materials. However, the technique's reliance on a narrow-bandwidth X-ray beam results in inherently low photon flux, which leads to long exposure times and limits its applicability for materials with a weak diffraction signal. Moreover, highly deformed materials with large intra-grain orientation spreads often require impractically long acquisition times, restricting real-time and high-throughput studies. These time constraints and limitations due to sample deformation also pose a major challenge for methods aiming to combine multiple Bragg reflections to reconstruct embedded tensor fields with DFXM \cite{Henningsson2025, Detlefs2025}. 
To address these limitations, pink-beam DFXM (pDFXM) was developed at the new beamline ID03 at the European Synchrotron Radiation Facility, ESRF \cite{isern2025esrf}. The pink beam is obtained by using a double multilayer monochromator (DMM) instead of the classical double crystal Si monochromator.  The pink beam was demonstrated to provide a 27-fold increase in intensities
\cite{yildirim2025pink}. 

However, the use of a broadened energy bandwidth may introduce chromatic aberration, deteriorating spatial resolution, similar to what has been observed in bright field X-ray microscopy \cite{Falch2016}. Likewise, the reciprocal-space resolution is impacted, altering strain sensitivity and dislocation contrast. The increased bandwidth also affects the effective angular selectivity, leading to potential trade-offs between improved photon efficiency and resolution degradation. Furthermore, the higher flux associated with pDFXM raises concerns regarding beam heating and, for some materials, beam-induced radiation damage.

In this study, we introduce a geometrical optics-based formalism for pDFXM to describe both direct and reciprocal space resolutions. Then, we present a systematic comparison of monochromatic and pink-beam DFXM, examining their respective resolution functions. We generalize existing geometrical optics theory \cite{Poulsen2017} to the pink beam case, providing closed expressions for key microscope characteristics. Next, we combine experimental measurements with numerical simulations to quantify the effects of beam divergence and energy bandwidth broadening on spatial resolution, angular selectivity, and weak beam contrast. 
Finally, we show the effect of pink-beam operations on the temperature rise of a specific specimen. Providing thermal decay times and impact on measurement protocols.

\section{Dark-field X-ray Microscopy geometry}

\begin{figure}
    \begin{center}
    \includegraphics[width=1\linewidth]{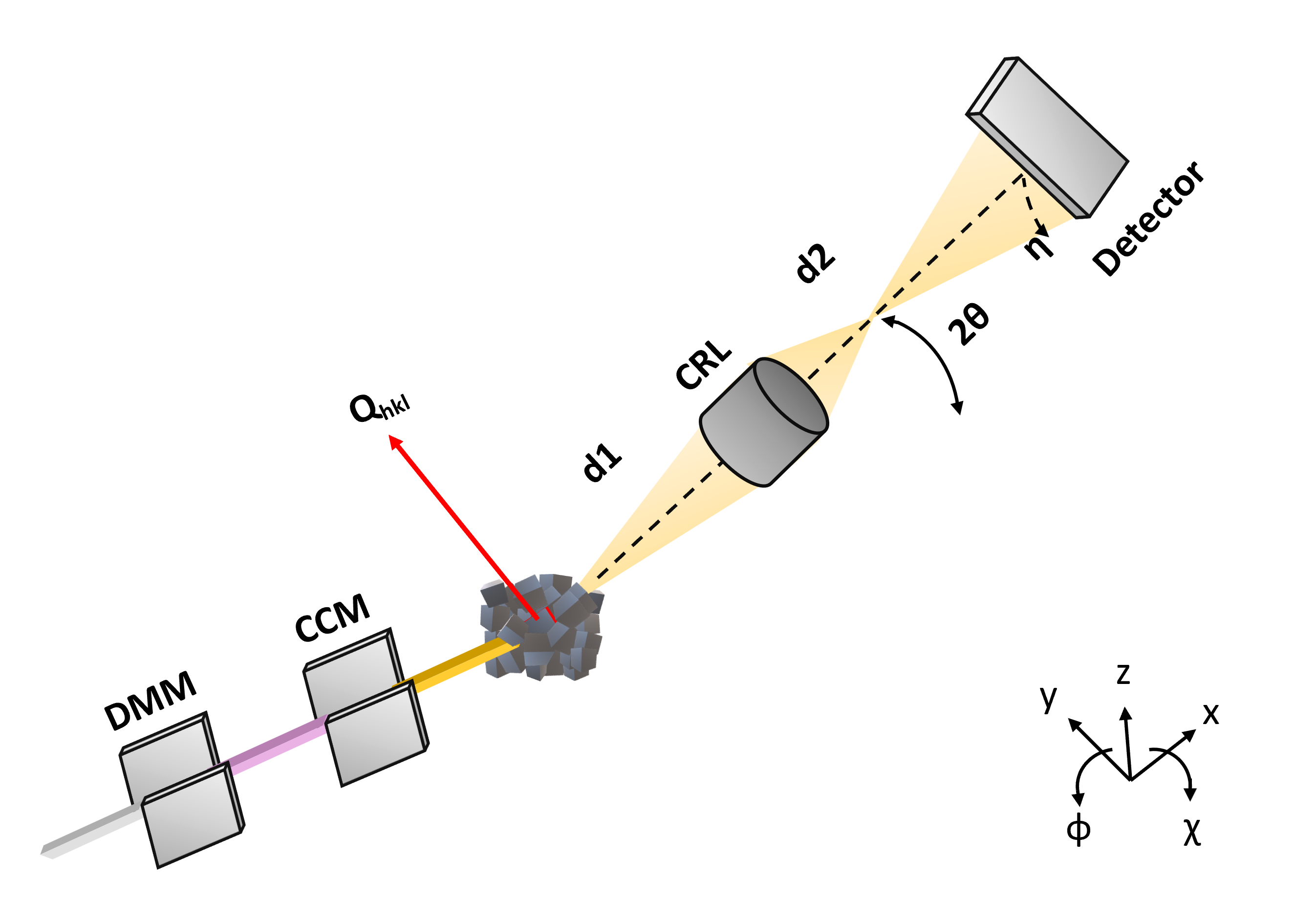}
    \caption{The DFXM set-up for the monochromatic illumination. The incident beam is shaped by the double multilayer monochromator (DMM), then by the channel cut monochromator (CCM). The objective is a Compound Refractive Lens (CRL). This is inserted in the line of the diffracted beam, characterized by angles $2\theta$ and $\eta$. The optical axis of the objective is shown as a dashed line with indications of the sample-to-entry-of-objective distance, $d_1$ and the exit-of-objective-to-detector distance, $d_2$.  }
    \label{fig_geometry}
    \end{center}
\end{figure}

Fig.~\ref{fig_geometry} outlines the geometry of DFXM. The energy of the incident beam is defined by either a channel-cut crystal monochromator --- in the following termed a monochromatic beam - or a double multilayer monochromator --- a pink beam. The energy bandwidth, $\epsilon = \Delta E/E$ with $E$ being the energy in eV, is approximately $10^{-4}$ and $10^{-2}$ in the respective cases. The incident beam can be focused vertically using a condenser to form a line beam, or left unfocused to produce a box beam. The incident beam has a divergence characterized by a full width at half maximum (FWHM) of $\Delta$$\zeta_v$ and $\Delta$$\zeta_h$ in the vertical and horizontal directions, respectively \cite{Poulsen2017}.

The sample can be either a single crystal or a polycrystalline aggregate. In both cases the goniometer movements are used to bring a specific grain or domain into the Laue conditions. The reflection of interest is indicated with a diffraction vector $\vec{Q}_{hkl}$. It is defined by the diffraction angle 2$\theta$ and an azimuthal angle $\eta$ \cite{Poulsen2017}. In the following we will assume a vertical scattering geometry: $\eta = 0$. 

Characteristic of DFXM, an objective is placed in the diffracted beam, such that the optical axis of the objective intersect with the pivoting point of the goniometer and a 2D detector. The objective magnifies diffraction contrast from the sample plane into an inverted 2D image on the detector. In this paper, the X-ray objective will be a Compound Refractive Lens  \citeaffixed{Snigirev1996}{CRL,} comprising $N$ identical parabolic lenslets, with a radius-of-curvature at the apex, $R$, and  a distance between neighboring lenslets of $T$. The distance from the sample to the CRL entrance is $d_1$, and the distance from the CRL exit to the detector is $d_2$. The total optical path length is $L = d_1 + NT + d_2$, where $NT$ is the combined thickness of the lens stack. The resulting magnification is $\mathcal{M}$.

The sample is mounted in a goniometer that enables rotations around multiple axes. Shown in Fig.~\ref{fig_geometry} are the two orthogonal tilts $\chi$ and $\phi$. Scanning $\phi$, $\chi$, and $2\theta$, a motion of CRL and detector together, corresponds to probing contrast by varying $\vec{Q}_{hkl}$ in three orthogonal directions.  These scans are known as rocking, rolling and axial strain scans, respectively. For a more complete geometrical description, see  \citeasnoun{Poulsen2017} and \citeasnoun{Poulsen2021}. The DFXM goniometer may include additional degrees of freedom. Details of the dedicated DFXM setup at beamline ID03 of the ESRF, used in this work, are given in \citeasnoun{isern2025esrf}.
  
\section{Geometrical optics formalism for pink beam}

A formalism for the main properties of the microscope was developed based on geometrical optics in \citeasnoun{Simons2017}, \citeasnoun{Poulsen2017} and \citeasnoun{Poulsen2018}. We here summarise key equations for the  monochromatic beam case and generalise these to the use of a pink beam. 

\subsection{Focal length, NA and Field-of-view}
\label{sec-fN_NA}

For brevity, we introduce the shorthand 
\begin{equation}
\varphi = \left(\frac{T}{f}\right)^{1/2}. 
\end{equation}
In monochromatic beam conditions, the focal length of one lenslet is  $f = R/(2\delta)$, where the refractive index decrement $\delta \propto 1/E^2$. For $N$ identical lenslets, the focal length becomes 
\begin{equation}
f_N = f \varphi \cot(N\varphi)
\end{equation}
 In a pink beam, $f_N$ varies with energy, leading to a relative spread of (thick or thin lens limit):
\begin{equation}
\Delta f_N /f_N \approx 2 \epsilon. \label{eq-spread_fN}
\end{equation}

In general, the average $f_N$ is linked to the magnification of the objective $\mathcal{M}$ by 
\begin{eqnarray}
 d_1 = & f_N \left[ \; 1 + 1/(\mathcal{M} \cos(N\varphi) ) \; \right] , \label{eq-d1}\\
d_2 = & f_N \left[ \; 1 + \mathcal{M}/\cos(N\varphi) \; \right] . \label{eq-d2}
\end{eqnarray}
It follows from Eqs. \ref{eq-spread_fN} and \ref{eq-d2} that the image plane is shifted along the optical axis by approximately $2 \, \epsilon \, d_2$ across the pink beam energies.  Stated differently, the change in magnification (for fixed $d_1$) becomes $\Delta  \mathcal{M}/\mathcal{M} \approx 2 \, \epsilon$. 

For the Numerical Aperture, NA, we shall consider \emph{the opaque lens case} where the effective aperture is dominated by attenuation in the parabolic part of the lens, and not the physical aperture.
For a monochromatic beam, the FWHM of the attenuation distribution in the on-axis case is \cite{Poulsen2017}:
\begin{eqnarray}
 \text{NA} = 2.35 \; \sigma_a = 2.35 \; \delta \frac{\mathcal{M}}{\mathcal{M}+1}  \sqrt{\frac{2N}{\mu R}},  \label{eq-NA}
\end{eqnarray}
with $\mu$ being the linear attenuation coefficient. The variation within an energy spread of 1\% is negligible; hence this equation is also valid for the pink beam case. The same is the case for the field-of-view (FOV). 

\subsection{Direct space resolution}

The \emph{depth of field}, $\sigma_{\text{depth}}$, for monochromatic beam DFXM is given by the combination of a waveoptics and a geometrical optics term, defined by the resolution in the imaging plane $y_d$ \cite{Poulsen2017}: 
\begin{equation}
2.35 \, \sigma_{\text{depth}} = \frac{\lambda}{(\text{NA})^2} + \frac{y_d}{\mathcal{M} \, \text{NA}} 
\end{equation}
This appears to be essentially constant over a 1 \% energy band. \emph{The chromatic aberration} is a concern for pink beam operation, as the focal length and magnification varies with energy as discussed in section \ref{sec-fN_NA} \cite{Poulsen2017}. This is expected to impact the data in several ways. Firstly, for line beam operation the vertical beam height will increase by approximately  $ 2Y_{\mathrm{con}} \, \epsilon $, where $2Y_{\mathrm{con}}$ is the effective physical aperture of the condenser - which may be defined by an upstream aperture. For typical operation at ID03,  $2Y_{\mathrm{con}} \approx 150 \mu$m and hence the beam-height becomes 1.5 $\mu$m for $\epsilon = 0.01$. 

Secondly, in relation to the objective, in \emph{bright field} there will be a lateral/transverse chromatic aberration --- smear in the radial direction, $r$ in the image, also known as the transverse chromatic aberration. This effect is described theoretically and demonstrated experimentally in \citeasnoun{Falch2016}. Here, it is also shown that one can mitigate the effect by focusing the beam from the condenser onto the plane of the objective.  However, 
this elegant solution and the underlying formalism do not apply to dark-field. Analytical expressions for this case are complicated, see  \citeasnoun{Simons2017}. In this paper we will instead provide numerical results based on DFXM ray-tracing, see section \ref{sub-Axel_simul}.

\subsection{Reciprocal space and strain resolution}

 The angular resolution of DFXM is essentially determined by properties of the incident beam and the diffracted beam (including the objective).  The  Darwin width intrinsic to the sample  can typically be neglected. 
 
Before entering into the formalism for the resolution function, it is useful to define three coordinate systems. Following \citeasnoun{Poulsen2021} we have:
\begin{itemize}
\item \emph{The laboratory system.}  This is described by the axes ($\hat{x}_{\ell}$, $\hat{y}_{\ell}$, $\hat{z}_{\ell}$) and is used to express the properties of the incident beam. 
\item \emph{The imaging system.} This is used to describe the diffracted beam. It has the optical axis of the objective as its x-axis, and is related to the laboratory system by a rotation around the y axis by $2\theta$.  The corresponding reciprocal space axis are defined as ($\hat{q}_{rock'}$, $\hat{q}_{roll}$, $\hat{q}_{2\theta}$). 
\item \emph{The crystal system.}  This is used to describe diffraction properties. It is defined as having the scattering vector $\vec{Q}$ along the z-direction and is described by the axes ($\hat{q}_{\text{rock}}$, $\hat{q}_{\text{roll}}$, $\hat{q}_{\parallel}$) in reciprocal space. It is related to the laboratory system by a rotation around the y-axis by $\theta$. 
\end{itemize}

The reciprocal space resolution at the intersection between the optical axis and the sample plane is described in the following. The projections of resolution function onto the three axes of \emph{the crystal system} are \cite{Poulsen2017}: 
\begin{eqnarray}
\Delta Q_{\text{rock}} = & \hspace{-2.5cm}\dfrac{\left| Q_0 \right|}{2} \left( \Delta \zeta_v^2 + \sigma_a^2 \right)^{1/2}, \label{eq-q_rock} \\
\Delta Q_{\text{roll}} = &\hspace{-2.1cm} \dfrac{\left| Q_0 \right|}{2 \sin(\theta_B)} \left( \Delta \zeta_h^2 + \sigma_a^2 \right)^{1/2}, \label{eq-q_roll} \\
\Delta Q_{\parallel} = & \hspace{0.2cm} \dfrac{\left| Q_0 \right|}{2} \left[ (2\epsilon)^2 + \cot^2(\theta_B) (\Delta \zeta_v^2 + \sigma_a^2) \right]^{1/2} ,  \label{eq-q_par}
\end{eqnarray}
with $\left| Q_0 \right|$ being the length of the nominal diffraction vector and $\theta_B$ being the Bragg angle. It is evident that a change in the energy bandwidth only affects $\Delta Q_{\parallel}$.  In the \emph{imaging coordinate system} on the other hand, due to the rotation of $\theta$ broadening will appear in both the rocking and the 2$\theta$ direction. Since the range of the reciprocal space probed is small, it is convenient to work with normalized diffraction vectors \citeaffixed{Poulsen2021}{strain units, }:
\begin{eqnarray}
q = & \dfrac{Q - Q_0}{\left| Q_0 \right|}. 
\label{q_normalised}
\end{eqnarray}

\subsection{Back focal plane} 

As for any classical microscope the objective in DFXM is associated with a Fourier plane, the Back Focal Plane, BFP. This is perpendicular to the optical axis. For completeness we discuss the properties in this plane in relation to an increased bandwidth.

Let the positions in the BFP be parameterized by $(y_B, z_B)$.  For a monochromatic and parallel beam the positions are linearly related to the angular deviations from the optical axis. Expressed in terms of the strain in the rolling and $2 \theta$ directions we have the relations \cite{Poulsen2018}:
\begin{eqnarray}
q_{roll} = & \dfrac{\cos (N\varphi)}{2 \sin(\theta) \; f_N} y_B, \\
q_{2\theta}= & \dfrac{\cos (N\varphi)}{2 \sin(\theta) \; f_N} z_B. 
\end{eqnarray}

These relations become more complicated when switching to a non-monochromatic and divergent pink beam \cite{Poulsen2017}: 

\begin{eqnarray}
q_{roll}= & \dfrac{1}{2 \sin(\theta)} \sqrt{\left(\frac{\cos (N\varphi)}{ f_N} y_B \right)^2 + \Delta\zeta_{h}^2}, \label{Eq_strain_roll_pink}\\
q_{2\theta} = & \dfrac{1}{2 \sin(\theta)} \sqrt{\left(\frac{\cos (N\varphi)}{f_N} z_B \right)^2 + \left(\cos(2 \theta)\Delta\zeta_{v} \right)^2 + \left(\sin(2 \theta)\epsilon \right)^2}. \notag \label{Eq_strain_2th_pink}\\
\quad
\label{q_res}
\end{eqnarray}

In the monochromatic beam case, the relatively poor axial strain resolution in DFXM can be improved by inserting an aperture or a knife-edge in the BFP. It appears from Eqs. \ref{Eq_strain_roll_pink}-\ref{Eq_strain_2th_pink} that in the pink beam case the strain resolution in the $2 \theta$ direction is ultimately limited by the energy bandwidth. Hence, the achievable strain resolutions in the $2 \theta$ direction are about $10^{-4}$ and $10^{-2}$, for monochromatic and pink beam, respectively. 

\section{Simulations}

Following previous work in \citeasnoun{Poulsen2017} and \citeasnoun{Borgi2024} we shall use Monte Carlo simulations to provide maps of the on-axis (normalized) reciprocal space resolution function for both pink and monochromatic beams. Next, we use a ray tracing simulation approach to provide insights into chromatic aberration.  For details of the experimental configurations used in the simulations - including the energy spectrum of the pink beam - and samples used, see Supplementary information section 1 and 5.

\subsection{Simulations of the reciprocal space resolution function}
\label{sub-simul_rec_space} 

The simulations were performed with the code presented in \citeasnoun{Borgi2024} and with $ 1 \times 10^{5}$ rays throughout. We assume that all properties of the microscope exhibit Gaussian distributions, defined by their rms widths.

\begin{figure}
    \begin{center}
    \includegraphics[width=0.9\linewidth]{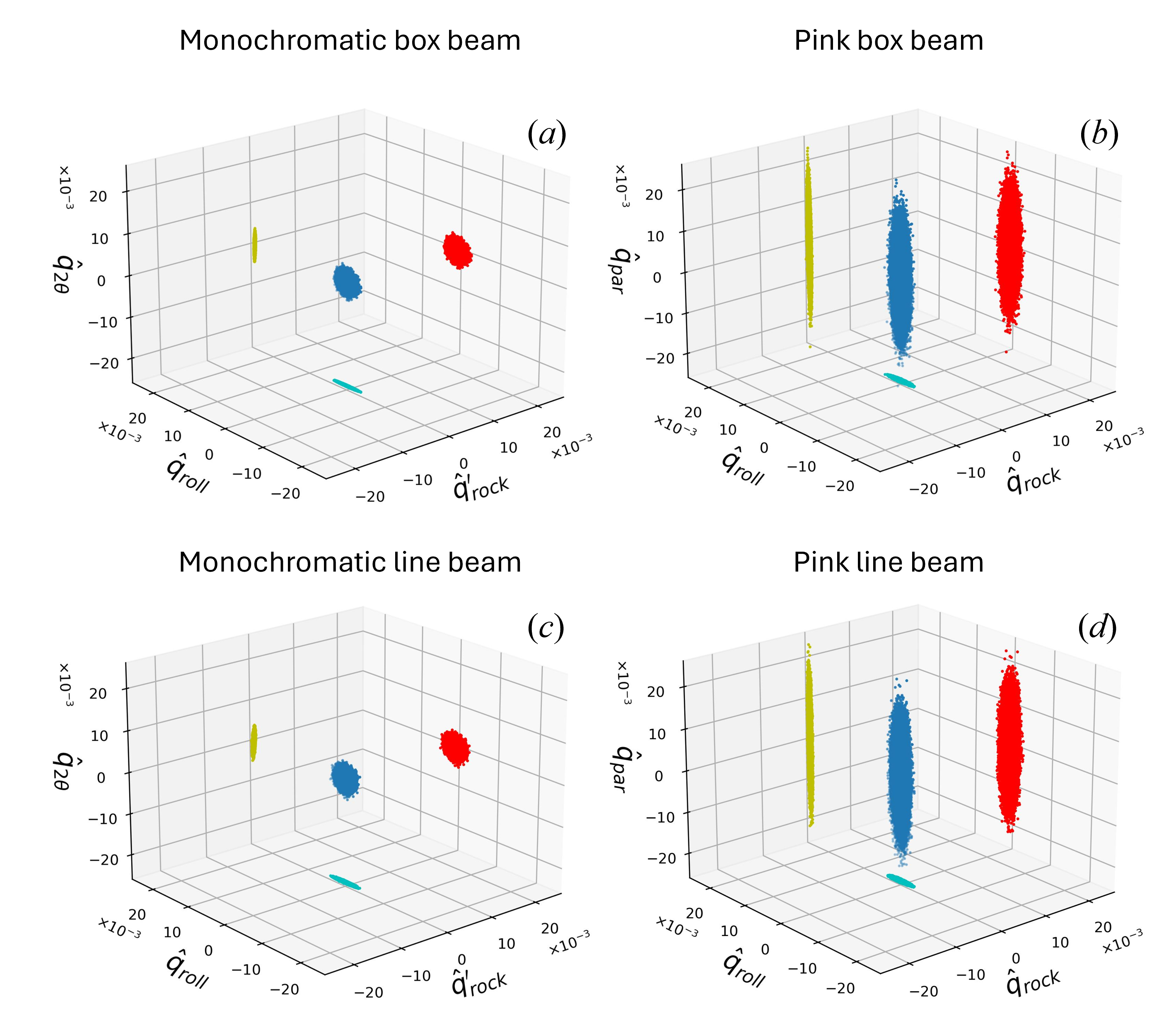}
    \caption{Reciprocal space resolution function for \emph{box beam (quasi-parallel illumination)} with monochromatic (a) and pink (b) beam, and \emph{line beam (condensed illumination)} for monochromatic (c) and pink (d) beam. In all cases, the reciprocal space is normalized, cf. Eq.~\ref{q_normalised}. The resolution function is displayed as a point cloud in 3D in  blue, while the red, yellow and cyan point clouds are projections onto 2D planes.}
    \label{simulations_box_line}
    \end{center}
\end{figure}

First, we consider \emph{the box beam case (quasi-parallel illumination)}. The resulting resolution function for the monochromatic beam is shown in the imaging systems in Fig. \ref{simulations_box_line} (a). The resolution function appears to be a disk with a diameter given by the NA of the objective and with the disk axis along $\Delta Q_{\text{rock}'}$. The fact that $ \Delta Q_{\text{rock}'} \ll \Delta Q_{\text{roll}} \simeq \Delta Q_{2\theta}$
has been key to the design of DFXM weak beam experiments. 

Components of the corresponding pink beam resolution function in the crystal system are shown Fig.\ref{simulations_box_line} (b). Essentially the figure represents a convolution of the monochromatic beam resolution function with a Gaussian in the direction of the diffraction vector; cf. Eq. \ref{eq-q_par}. This has several consequences for pDFXM: 
\begin{itemize}
\item \emph{Deterioration of the axial strain resolution}. Spanning a range in Q$_{\parallel}$ of $4 \times 10^{-2}$ makes axial strain determination infeasible. On the other hand, pink beam operation implies that one typically will be able to ensure integration over the entire strain profile, and hence may speed up experiments that depend on sampling the integrated intensity.
\item \emph{Deterioration of the rock' resolution.}  Weak beam operation has benefited from the superior resolution in the direction perpendicular to the disk. 
In the pink beam mode the thinnest direction in reciprocal space is ten times thicker.
\item \emph{Choice of coordinate system}. The directions of the principal axes change. While the imaging system is the natural choice for monochromatic beam operation, the crystal coordinate system is more relevant for pink beam operation. This choice of coordinate system is conceptually more intuitive for applications.  
\end{itemize}


Next, we consider \emph{the line beam case (condensed illumination)}. The resulting resolution functions are shown in Figs. \ref{simulations_box_line} (c) and (d). While the increased divergence has some effect in the monochromatic beam case, this contribution is completely negligible in the pink beam case. The latter can be an advantage: it facilitates direct quantitative comparisons between line and box beam mapping.

\subsection{Simulations of chromatic aberration}\label{sub-Axel_simul}
The broadened energy bandwidth in pDFXM does not only impact reciprocal space resolution, as investigated in the previous section, but will also deteriorate spatial resolution by blurring detector images in direct space. The effect is known as \textit{chromatic aberrations}, and is well known from bright field X-ray microscopy \cite{Falch2016}. The corresponding effect in dark-field, has, to the best of our knowledge, never been studied before. Simulating these effect requires to trace X-rays through the 3D objective lens-stack as a function of wavelength. Unfortunately, previous geometrical optics work in DFXM can not account for such ray-trajectories. In frameworks such as \citeasnoun{Poulsen2017} , \citeasnoun{Borgi2024} and \citeasnoun{Henningsson2025} each point in the sample plane is approximated to deposit all diffracted photons at a corresponding, fixed detector pixel, neglecting depth-of-focus and vignetting effects. While this allows for fast simulations, where the reciprocal resolution function can be used as a look-up table for diffracted intensities, it does not allow to simulate imaging artifacts. To progress, we have develop a new framework that traces rays in 3D through the objective lens-stack. 

After a photon has scattered of the sample, we use Snells's law to solve a recursive set of non-linear equations that trace the ray path through the beryllium lens-stack. At each lens-let, refraction from the parabolic entry and exit surface is computed, and the path length through the lens is used to attenuate the ray intensity using Beer's law. Both of these effects are taken to be functions of the randomly sampled wavelength of the ray such that each traced ray perceives a unique refractive decrement, \(\delta=\delta(E)\), and attenuation coefficient, \(\mu=\mu(E)\). A detailed mathematical derivation of our framework is provided in SI, section 5.

Importantly, in our new framework, photons that are sampled from the direct beam undergo scattering following the Laue equations. The resulting diffracted beam statistics is therefore a modulation of the direct beam via an interaction with the sample lattice. This makes dark-field chromatic aberrations inherently \textit{sample dependent}. This is an important conclusion for pDFXM that prohibits us from generalizing a single simulation across experiments. Nevertheless, in the following we present simulations of what can be considered to be the canonical case; undeformed single-crystals placed at the center of focus in the sample plane. By using experimentally observed statistics for the direct beam divergence and bandwidth these simulations quantifies how severe chromatic artifacts are for practical imaging scenarios. All driving distributions are taken to be Gaussian an independent. Four distinct scenarios are examined:

\textbf{A} -- \emph{ Parallel and Monochromatic Beam}: Vertical and horizontal divergence of the direct beam were taken as \(\xi_v=\xi_h=0.01\) mrad (FWHM). The energy bandwidth was taken as 0.001146 keV (standard deviation). A $BiFeO_{3}$-002 reflection is considered. 

\textbf{B} -- \emph{ Condensed and Monochromatic Beam}
Vertical and horizontal divergence of the direct beam was taken as \(\xi_v=0.53\) and \(\xi_h=0.01\) mrad (FWHM) respectively. The energy bandwidth was taken as 0.001146 keV (standard deviation). An Si--220 reflection is considered. 

\textbf{C} -- \emph{ Parallel and Pink Beam}: Vertical and horizontal divergence of the direct beam was taken \(\xi_v=\xi_h=0.01\)mrad (FWHM). The energy bandwidth was 0.10123 keV (standard deviation). A $BiFeO_{3}$-002 reflection is considered. 

\textbf{D} -- \emph{ Condensed and Pink Beam} Vertical and horizontal divergence of the direct beam was taken as \(\xi_v=0.53\) and \(\xi_h=0.01\) mrad (FWHM) respectively. The energy bandwidth was 0.10123 keV (standard deviation). An Si--220 reflection is considered.

The mean energy was set to 19.1 keV and the sample to CRL and CRL to detector distances were adjusted accordingly. For parallel beam, the placement was taken to yield a 11.26 times magnification from the CRL. For the condensed cases, the CRL magnification was fixed to 15.77. The CRL featured 87 parabolic Berylium lens-lets with \(R=50\mu\)m, \(T=1.6\)mm and web-thickness 30 \(\mu\)m. In all scenarios, the raw detector pixel size was \(6.5\mu\)m and the optical magnification of the detector was \(10 \times\).

\begin{figure}
  \centering
    \caption{Simulated diffraction response from ring-shaped, undeformed single crystals under different beam conditions. When the beam is both pink and condensed (D1 and D2), strong chromatic aberrations appear. In A1--D1, the rings are centered on the optical axis, while in A2--D2, they diffract toward the detector edge. Comparing D1 and D2 shows that chromatic aberrations depend on the offset from the optical axis. Chromatic aberrations are much stronger in the vertical scattering plane (detector \(z\)) than in the transverse plane (detector \(y\)). Cartesian line profiles across the ring edges are shown in E for all four pink-beam scenarios (C1--D2). Chromatic aberrations in D1--D2 are seen to cause a \(1~\mu\text{m}\) blurring , while the pink-parallel beam cases (C1--C2) show no blurring.}
  \label{fig:pinkdiscs}
  \makebox[\textwidth][c]{\includegraphics[width=0.99\textwidth]{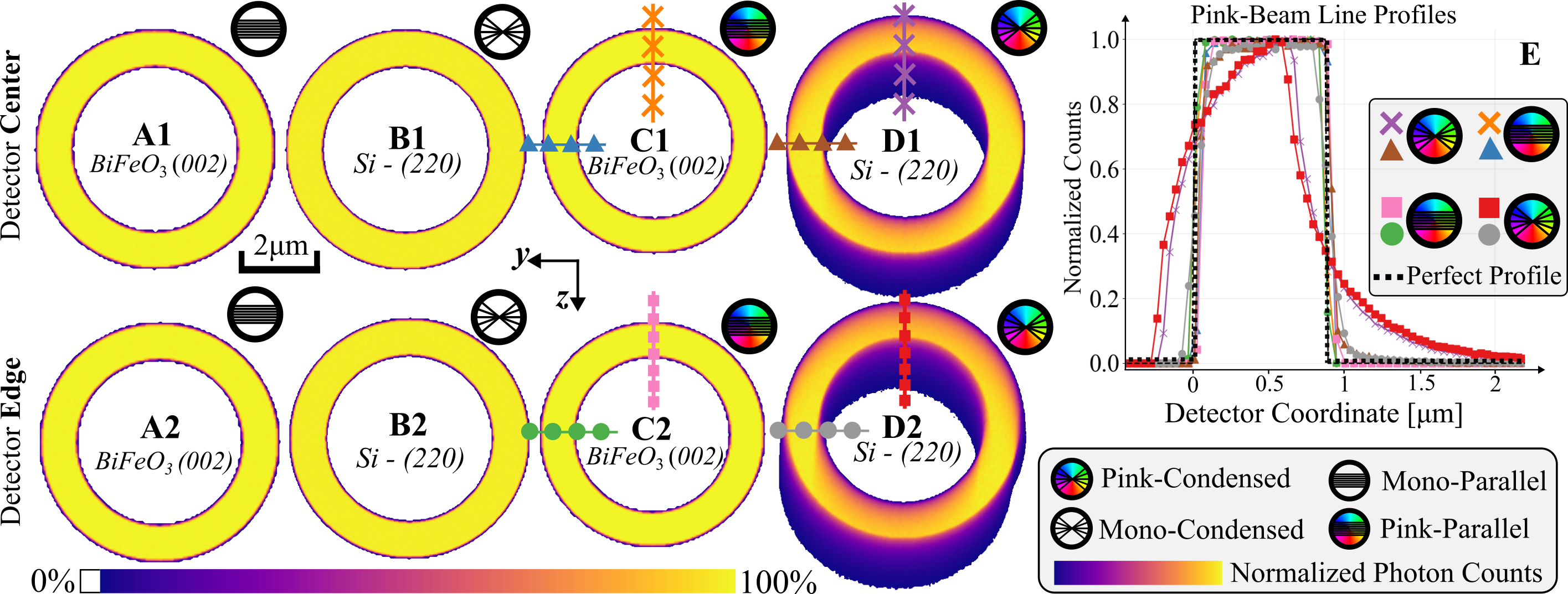}}
\end{figure}

A grid of 81 ring-shaped single crystals ($BiFeO_{3}$-002 and Si-220 reflections) were placed in the sample plane and Monte-Carlo methods were used to sample rays from the beam statistics defined though A-D. The resulting detector image for two selected rings and all beam scenarios (A-D) are presented in Figure \ref{fig:pinkdiscs}. In A1--D1, the rings centered on the optical axis have been selected, while in A2--D2, off-axis rings are shown (diffracting towards the detector edge). Each individual ring corresponds to a total of 84 million rays individually traced through the CRL lens-stack.

Comparing D1 and D2 shows that chromatic aberrations depend on the offset from the optical axis. Chromatic aberrations are observed to be stronger in the vertical scattering plane (detector \(z\)) than in the transverse plane (detector \(y\)). Cartesian line profiles across the ring edges are shown in Figure \ref{fig:pinkdiscs}E for pink-beam scenarios (C1--D2). The chromatic aberrations in D1--D2 are seen to cause a \(1~\mu\text{m}\) blurring, while the pink-parallel beam cases (C1--C2) show no blurring.

The lack of a chromatic aberrations in Figures \ref{fig:pinkdiscs}A1-C2 is explained by the sample acting as a monochromator, filtering the direct beam though the Bragg condition. Consequently, only when the direct beam features \textit{both} a high vertical divergence \textit{and} a broad energy band does chromatic aberrations appear for undeformed crystals (D1-D2). In this special case, a divergent ray can diffract off the sample by being paired with an off-set wavelength such that the combination again fulfills scattering conditions. Many such interactions combine into a non-negligible energy broadening being present also in the diffracted beam. Notably, the resulting statistics of the diffracted beam are \textit{non-Gaussian}. Consequently, dark-field aberrations do not follow the patterns predicted in bright-field microscopy \cite{Falch2016}. 

The fact that the sample modulates the statistics of the diffracted beam, and hence the resulting aberrations, has important implications for crystals featuring \emph{a non-perfect lattice}. 
\onecolumn
\twocolumn
In such cases, the local mosaic spread will result in diffracted beam statistics featuring significant divergence and bandwidth -- \emph{even if the incident beam is parallel}. In other words, the divergence of the incident pink-beam in D1-D2 can be effectively replaced by a mosaic spread. It is therefore possible that for samples with moderate to large mosaic spreads the chromatic aberration can be worse than illustrated in Figure \ref{fig:pinkdiscs}E. As an illustration, for crystals that partition into several near-perfect cell-domains, separated by low-angle orientation boundaries, one may expect cell boundaries to exhibit strong chromatic aberrations, while cell interiors will be aberration free.

\subsection{Strong and weak beam operation}
\label{sub-sub-simul_direct_space}

The larger reciprocal space resolution function - as derived in the section~\ref{sub-simul_rec_space}- affects the optimal conditions for weak beam operation.  This is discussed in \citeasnoun{Borgi2024}.  Here DFXM forward projections are made of a wall of identical edge dislocations. Images are generated as function of the rocking angle, $\phi$, showing little or no dislocation contrast for $\phi$ close to 0 (strong beam). The contrast gradually improves with increasing $\phi$ (weak beam) while intensities are decreasing as the volume of the sample that gives rise to diffraction (close to the core of the dislocations) become smaller. Comparing such results for different energy bandwidths it was found that the strong beam region extends to larger $\mid \!\! \phi \!\!  \mid$ values for larger $\epsilon$ and that one consequently has to go to larger $\mid \!\! \phi \!\! \mid$ values for optimal conditions for weak beam imaging. 
Larger values implies higher strain values and hence diffraction from regions closer to the core of dislocations. Hence, there are two competing effects on the S/N of dislocation imaging: larger $\epsilon$ implies a lager incident flux, while at the same time leading to the diffraction volumes being smaller. The simulations in \citeasnoun{Borgi2024} indicate that the increased incident flux - for the case simulated - significantly improve the counting statistics with increased energy bandwidth. 

\section {Experimental results}
We conducted experiments for comparison with the analytical and numerical results presented above. Sample and configuration details are given in the Supplementary Information section 1, 2, and 3.
\subsection{Reciprocal space resolution function} 
To probe the rocking and rolling components, a large diamond single crystal was used as the sample, diffracting the (111) reflection. 

\begin{figure}
    \begin{center}
    \includegraphics[width=0.9\linewidth]{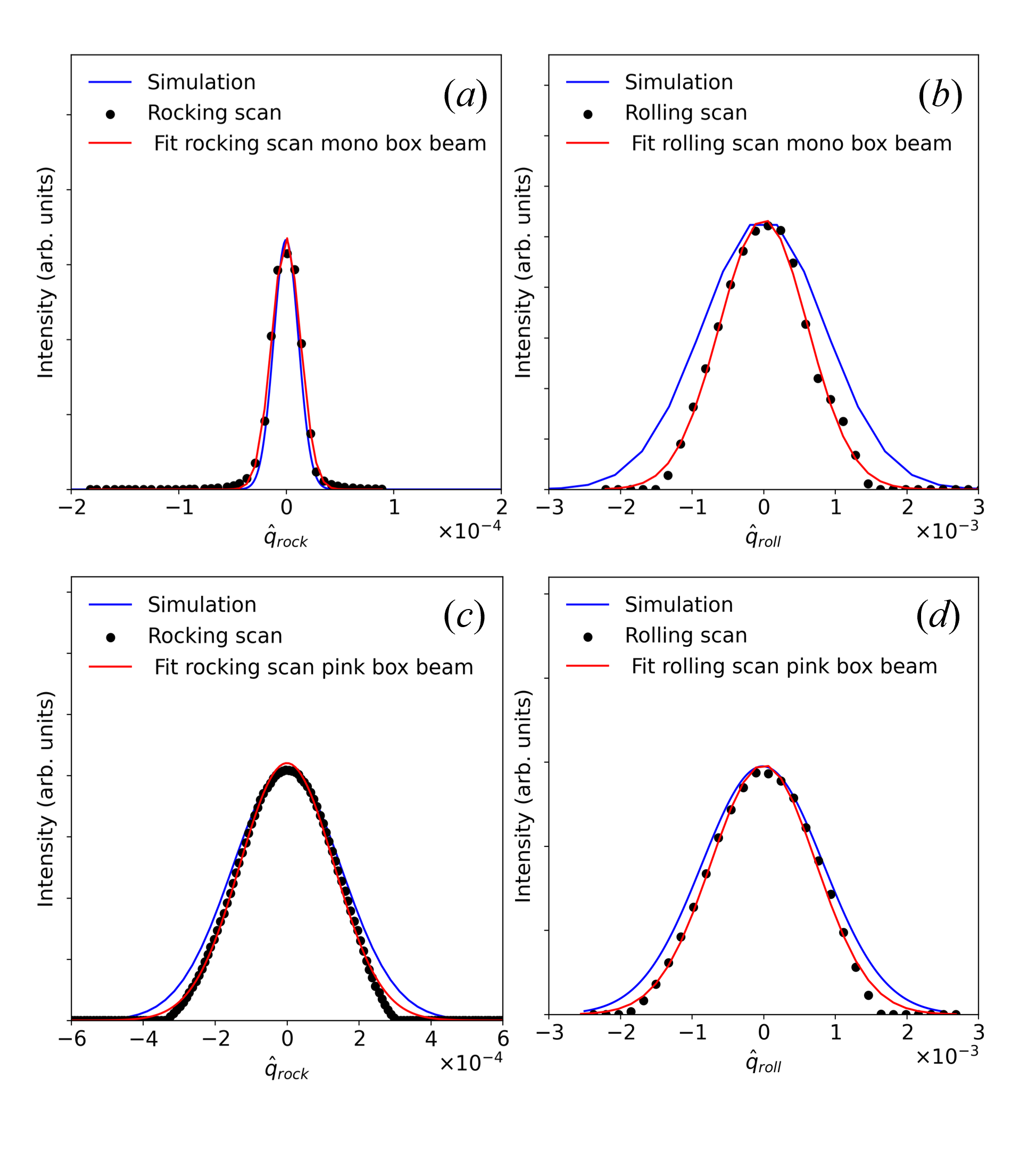}
    \caption{Comparison of normalized experimental data (dots) and simulations (blue curves) of the rocking and rolling direction components of the reciprocal space resolution function for the \emph{box beam} case. Also shown are fits of the experimental data to Gaussian distributions (red lines). a) Resolution function in $q_{\text{rock}'}$ for monochromatic beam. b) Resolution function in $q_{\text{roll}}$ for monochromatic beam. c) Resolution function in $q_{\text{rock}'}$ for pink beam. d) Resolution function in $q_{\text{roll}}$ for pink beam.}
    \label{box_beam_fig}
    \end{center}
\end{figure}

In the \emph{box beam} case (Fig.~\ref{box_beam_fig}), the simulations show excellent agreement with the experimental data for both directions and for both monochromatic and pink-beam illumination. In the rolling direction, the simulations overestimate the angular spread of 0.00057 strain units; however, as shown by \citeasnoun{Poulsen2017}, this effect arises from the scattering vector moving off the rocking curve. The corresponding FWHM values are listed in Table~\ref{tab:fwhm_boxbeam}.

\begin{figure}
    \begin{center}
    \includegraphics[width=0.9\linewidth]{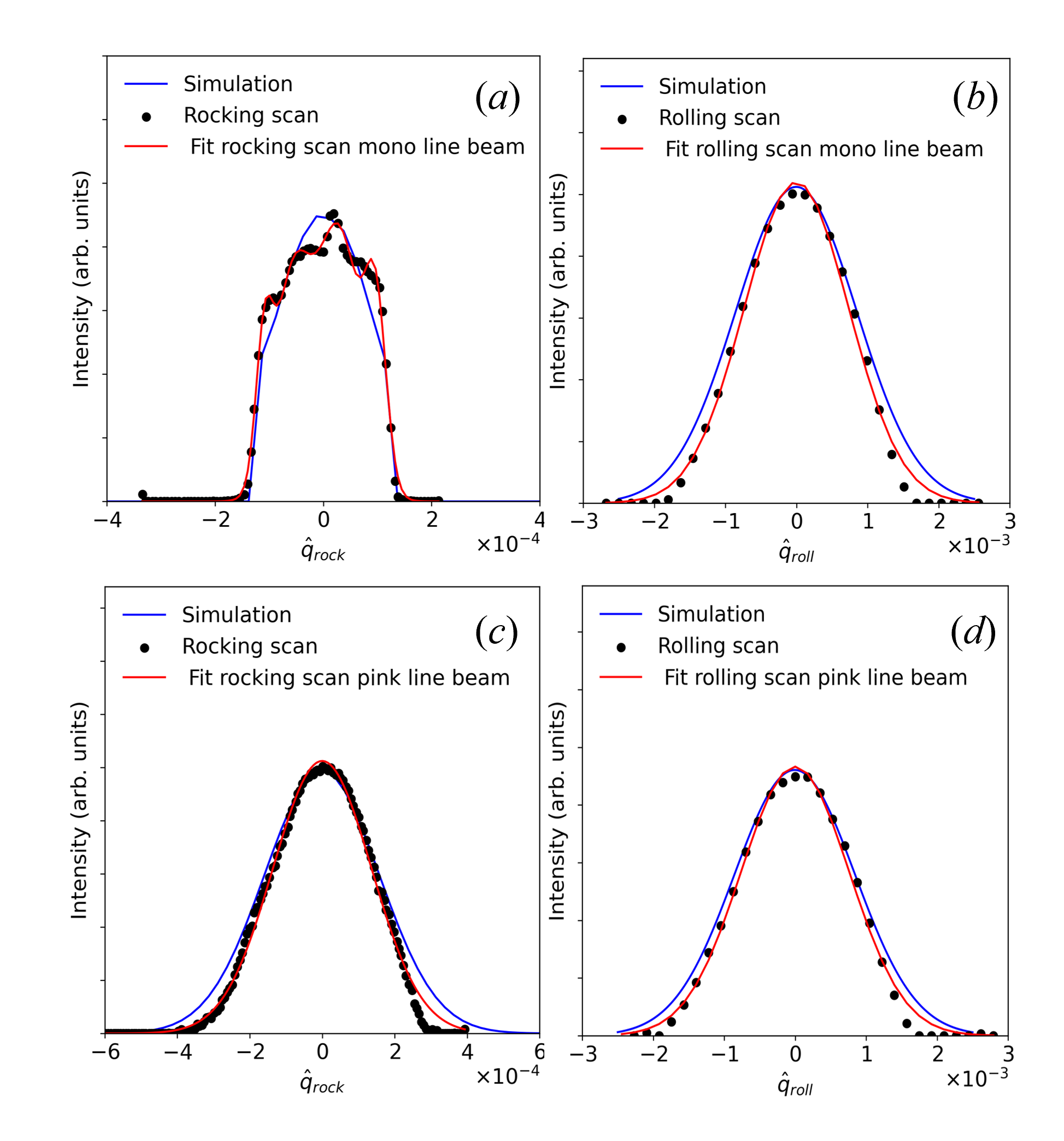}
    \caption{Comparison of normalized experimental data (dots) and simulations (blue curves) of the rocking and rolling direction components of reciprocal space resolution function for the \emph{line beam} case. Also shown are fits of the experimental data to Gaussian distributions (red lines). a) Resolution function in $q_{\text{rock}'}$ for monochromatic beam. b) Resolution function in $q_{\text{roll}}$ for monochromatic beam. c) Resolution function in $q_{\text{rock}'}$ for pink beam. d) Resolution function in $q_{\text{roll}}$ for pink beam.}
    \label{line_beam_fig}
    \end{center}
\end{figure}

In the \emph{line beam} case (Fig.~\ref{line_beam_fig}), there is also good agreement between the simulated and experimental curves. For the monochromatic beam, the angular resolution function in the rocking direction (Fig.~\ref{line_beam_fig}a) is truncated by a slit placed upstream of the condenser. Moreover, the $\xi_\mathrm{v}$ distribution was non-uniform, and both effects were accounted for in the simulations. In this case, the experimental curve was fitted with a multi-Gaussian model. The corresponding pink-beam resolution function (Fig.~\ref{line_beam_fig}b) appears smoother due to convolution with the energy bandwidth, though a slight truncation of the tails remains. The resulting FWHM values are listed in Table~\ref{tab:fwhm_linebeam}.  

\begin{figure}
    \begin{center}
    \includegraphics[width=1\linewidth]{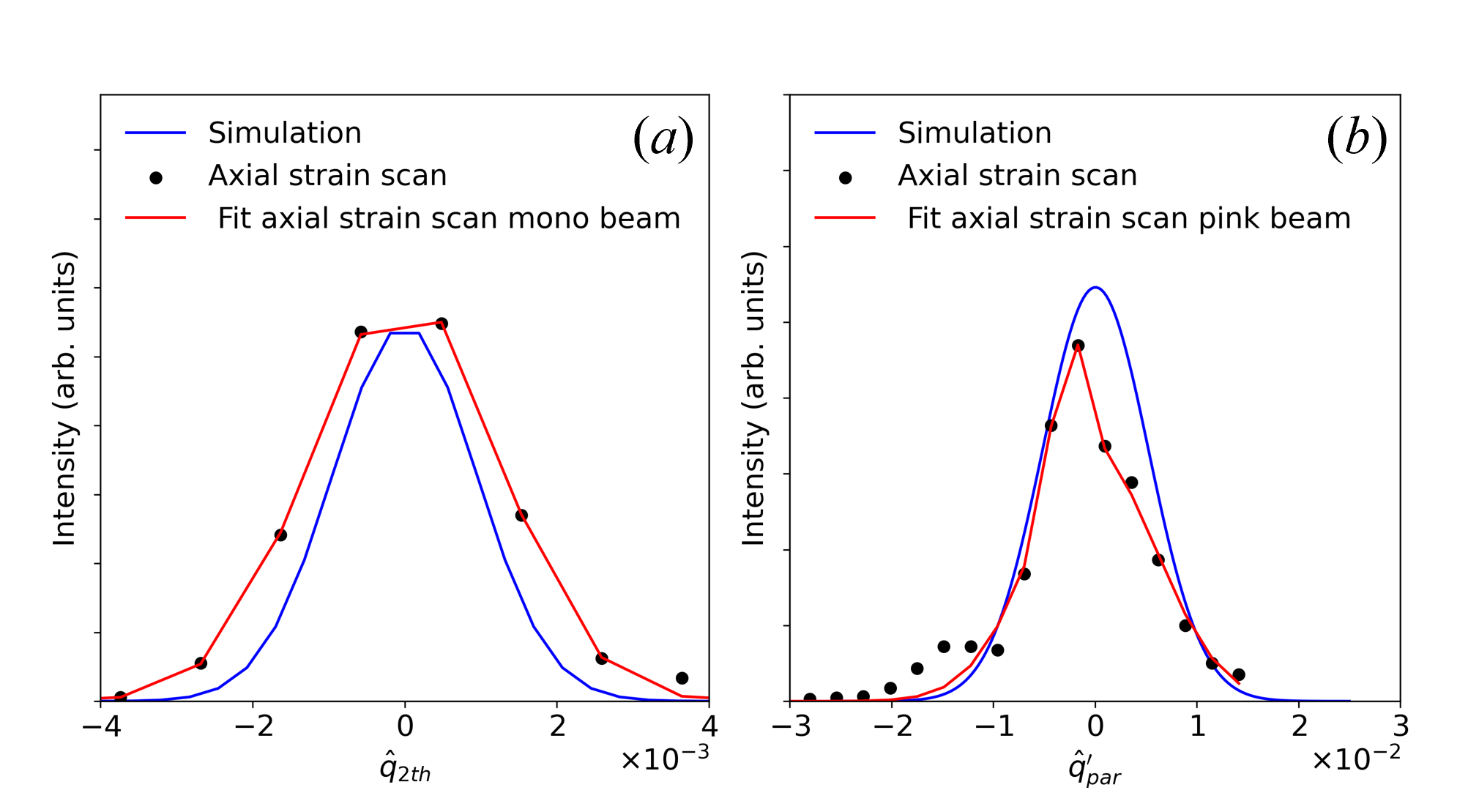}
    \caption{Comparison of normalized experimental data (dots) and simulations (blue curves) of the axial strain resolution function for the box beam case. Also shown are fits (red curves) of the experimental data. a) Resolution function in $q_{2\theta}$ for monochromatic beam. b) Resolution function in $q_{\parallel}$ for pink beam.}
    \label{strain_fig}
    \end{center}
\end{figure}

Corresponding data for the axial strain direction are shown in Fig.~\ref{strain_fig}. The experimental sampling is coarser in this case. Within the statistical uncertainty, a Gaussian model (blue curve) provides a reasonable approximation of the monochromatic beam data (Fig.~\ref{strain_fig}a). The pink beam curve, however, deviates from a Gaussian shape and can be approximated by the asymmetric energy profile (see Supplementary Information). Accordingly, the monochromatic beam data were fitted using a single Gaussian model, while a multi-Gaussian fit was applied to the pink-beam data. The fitted FWHM values from both experimental and simulated datasets are listed in Table~\ref{tab:fwhm_boxbeam}. The results overall confirm the expected reduction in resolution for the pink beam.

\subsection{Chromatic aberration}
\label{experiment_direct_space}

The simulations in Section~\ref{sub-Axel_simul} indicate that the extent of blurring in direct space depends strongly on both the sample type and its local deformation state. At present, however, there is no established reference sample for DFXM comparable to the resolution target used in bright-field imaging. Consequently, a direct one-to-one validation of the theoretical predictions is not possible. Thus, to qualitatively evaluate the effects of chromatic aberration, we conducted two representative experiments using quasi-parallel (box beam) and condensed (line beam) illumination. The local spatial resolution was quantified using a Fourier transform-based analysis, described in detail in the Supporting Information, section 2, along with the experimental configurations.

\begin{figure}
    \begin{center}
    \includegraphics[width=1\linewidth]{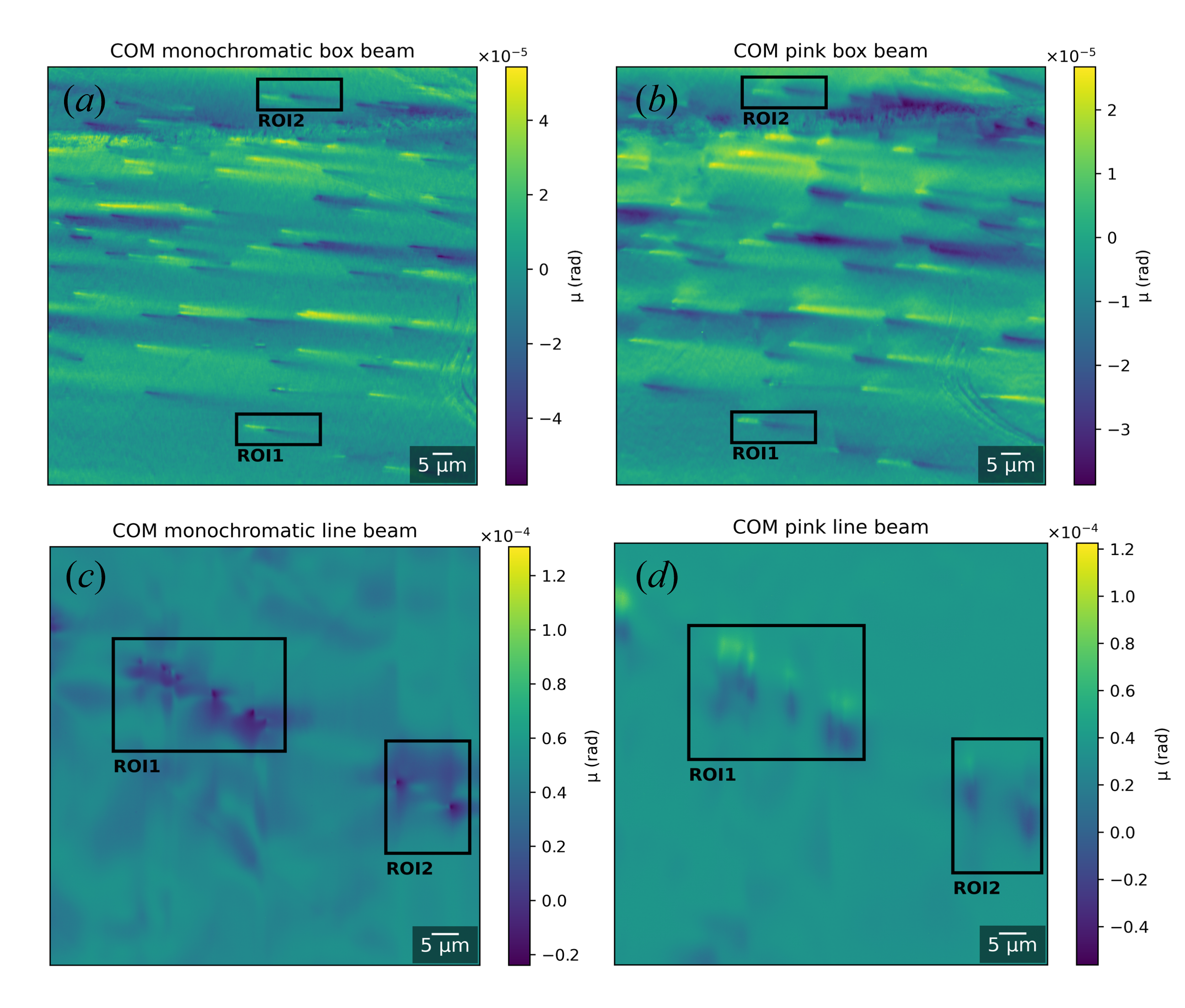}
    \caption{COM maps of the rocking curve measurements done on a) the $BiFeO_{3}$ sample with monochromatic box beam, b) the $BiFeO_{3}$ sample with pink box beam, c) the Si wafer sample with monochromatic line beam, d) the Si wafer sample with pink line beam. The ROIs used for the quantification of the resolution are indicated with black rectangles.}
    \label{COM_box_line_beam}
    \end{center}
\end{figure}

\emph{Box beam mode}. We have produced center-of-mass (COM) maps from rocking scans of the thin-foil $BiFeO_{3}$ sample using the Darfix software \cite{Ferrer2023DarfixMicroscopy}. The data exhibited distinct and well-defined dislocations. We used these features to quantify the spatial resolution. Two ROIs, both including single dislocations, were selected from the COM map. The resulting images, shown in Fig.\ref{COM_box_line_beam}a and b, for monochromatic and pink beam, respectively, exhibited a lateral shift. Comparing the spatial resolutions obtained, we have found that the broader bandwidth of the pink beam radiation only introduces a slight blur in the direct space image, leading to a decrease of resolution on the order of a few hundred nm. The values of resolution together with their standard deviations for the ROIs are listed in Table\ref{tab:box_line_resolution}.

\emph{Line beam mode}. In this case, we used a bulk Si wafer sample containing well-defined dislocations generated by indentation and annealing treatments. Again, we integrated the voxel intensities in a rocking scan. We focused on two ROIs at the center of the sample, presenting a network of well-known and previously studied dislocations \cite{kabukcuoglu20213d}. The COM maps shown in Fig.\ref{COM_box_line_beam}c-d were treated using the same protocol as for the box beam case. In this case, the decrease in resolution going from monochromatic to pink beam is on the order of 1-2 $\mu$m, higher than that of the box beam case. The results and standard deviations of the resolution quantification for both ROIs are listed in Table\ref{tab:box_line_resolution}.

\subsection {Weak beam contrast}
Mapping crystalline defects such as dislocations with DFXM has relied on the weak beam contrast \cite{Jakobsen2019,Dresselhaus2021,Yildirim2023} enabled by the superior angular resolution in the rocking direction. \citeasnoun{Borgi2024} simulated the effect of increasing bandwidth on the weak beam contrast. They generated images of a wall of edge dislocations with bandwidths increasing from $\epsilon = 10^{-4}$ to $10^{-2}$, using a fixed incident divergence, corresponding to the use of the condenser. They also discussed the contrast as function of $\phi$ offset, and proposed an 80 \% integrated intensity reduction as the optimal.

\begin{figure}
  \centering
    \caption{Weak beam contrast for studying dislocations with line beam illumination in monochromatic and pink radiation cases. In this case the pink beam was attenuated by a 50 $\mu$m Cu foil.  a) Rocking curves obtained by summing intensities in each detector pixel over all images in the DFXM rocking scan. The intensity of the rocking curves was normalized to 1 to facilitate direct comparison. The blue curve was acquired with the monochromatic beam, and the orange curve with the pink beam. Weak beam contrast images: b) monochromatic beam at a rocking angle corresponding to  $20\%$ of the maximum intensity (blue dashed line in a)).  c) Pink beam at at rocking angle corresponding to  HWHM (innermost orange dashed line in a)) d) Pink beam at at rocking angle corresponding to $20\%$ of the maximum intensity(outermost orange dashed line in a)).}
 \label{fig:weakbeam}
  \makebox[\textwidth][c]{\includegraphics[width=0.99\textwidth]{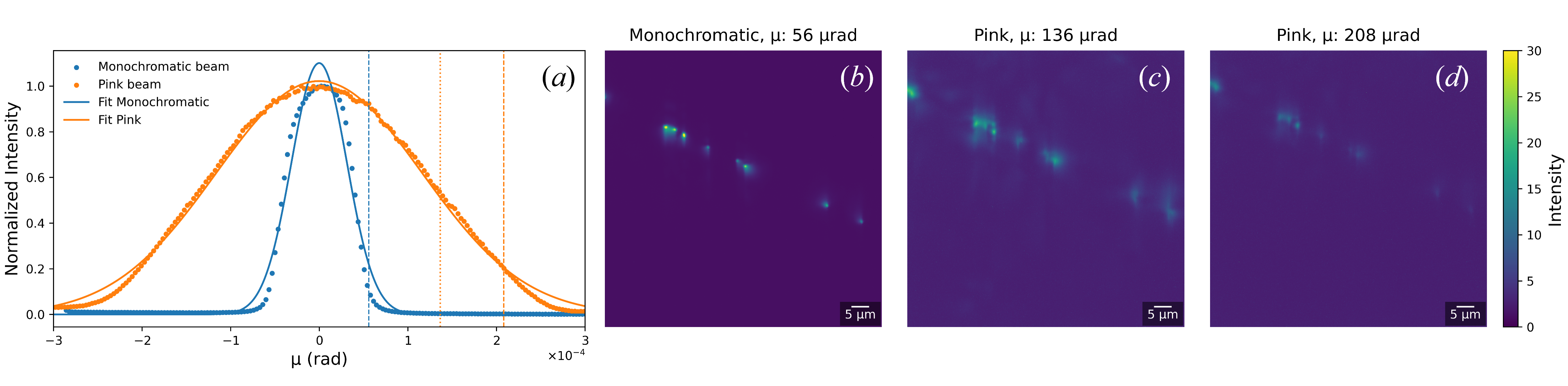}}
\end{figure}

In an attempt to establish comparable experimental data we studied a chain of dislocations in the Si wafer sample,  using the same datasets as presented above in connection with the direct space resolution quantification.  The results for the line beam case are shown in Fig.~\ref{fig:weakbeam}. As discussed above the rocking curve from the pink beam illumination is in this case substantially wider than the one from the monochromatic beam. 

Comparing the results in  Figs.~\ref{fig:weakbeam} b), c) and d) we conclude that the "80 \%" rule does not apply to pink beam. It is possible to acquire images without any (visible) effects of the strong beam at rocking angles corresponding to the HWHM of the rocking curve, This in conjunction with the high intensity in pink beam implies that an improved S/N can be obtained in pink beam mode, at the expense of a deterioration in spatial resolution. 

Corresponding results for the box beam case are provided in the Supplementary Information section 3.

\begin{table}[1]
\centering
\caption{Widths (FWHM) of the experimental and simulated strain resolution function  in the box-beam case}
\label{tab:fwhm_boxbeam}
\begin{tabular}{c|cc|cc}
\toprule
\textbf{Radiation} & \textbf{Exp. data} & \textbf{FWHM } & \textbf{Simulation} & \textbf{FWHM } \\
\midrule
mono & rocking & 0.00003 & rocking & 0.00003 \\
mono & rolling & 0.00145 & rolling & 0.00202 \\
mono & axial strain & 0.00288 & axial strain & 0.00222\\
\midrule
pink & rocking & 0.00031 & rocking & 0.00035\\
pink & rolling & 0.00173 & rolling & 0.00199 \\
pink & axial strain & 0.01026 & axial strain & 0.01227      \\
\bottomrule
\end{tabular}
\end{table}

\begin{table}[2]
\centering
\caption{Widths (FWHM) of the experimental and simulated strain resolution function   in the line-beam case}
\label{tab:fwhm_linebeam}
\begin{tabular}{c|cc|cc}
\toprule
\textbf{Radiation} & \textbf{Exp. data} & \textbf{FWHM } & \textbf{Simulation} & \textbf{FWHM } \\
\midrule
mono & rocking & 0.00026 & rocking & 0.00023 \\
mono & rolling & 0.00174 & rolling & 0.00201 \\
\midrule
pink & rocking & 0.00032 & rocking & 0.00037\\
pink & rolling & 0.00180 & rolling & 0.00200 \\
\bottomrule
\end{tabular}
\end{table}

\onecolumn
\twocolumn

\begin{table}
\centering
\caption{Quantified dark-field direct-space resolution, with its SD, in the box and line beam illumination modes.}
\label{tab:box_line_resolution}
\resizebox{\columnwidth}{!}{%
\begin{tabular}{lccc}
\toprule
\textbf{Illumination} & \textbf{Radiation} & \textbf{ROI} & \textbf{Resolution (nm)} \\
\midrule
\multirow{4}{*}{Box beam} 
 & \multirow{2}{*}{Monochromatic} & 1 & 529 $\pm$ 19\\
 &                                 & 2 & 821 $\pm$ 58 \\
 & \multirow{2}{*}{Pink}           & 1 & 799 $\pm$ 80\\
 &                                 & 2 & 959 $\pm$ 81 \\
\midrule
\multirow{4}{*}{Line beam} 
 & \multirow{2}{*}{Monochromatic} & 1 & 1984 $\pm$ 284 \\
 &                                 & 2 & 1600 $\pm$ 443\\
 & \multirow{2}{*}{Pink}          & 1 & 3266 $\pm$ 149 \\
 &                                 & 2 & 4127 $\pm$ 106\\
\bottomrule
\end{tabular}%
}
\end{table}

\section {Effects of beam heating}
Imaging at fourth-generation synchrotrons is known to cause sample heating. Regardless of S/N, pink beam operation intensifies heating, as a larger fraction of the direct beam is not diffracted by the 
sample and instead contributes to absorption. The extent and structural impact of heating depends on the
specimen and setup: in metals, dislocations become mobile before grain boundaries (recovery precedes 
recrystallization and growth), while systems near phase transitions are especially sensitive. DFXM 
scans are also non-isothermal, as the beam shutter cycles during acquisition, causing transient thermal 
stresses that shift diffraction angles. As a result, measurements can depend on scan timing, motor 
speeds and waiting periods.

\begin{figure}
    \begin{center}
    \includegraphics[width=1\linewidth]{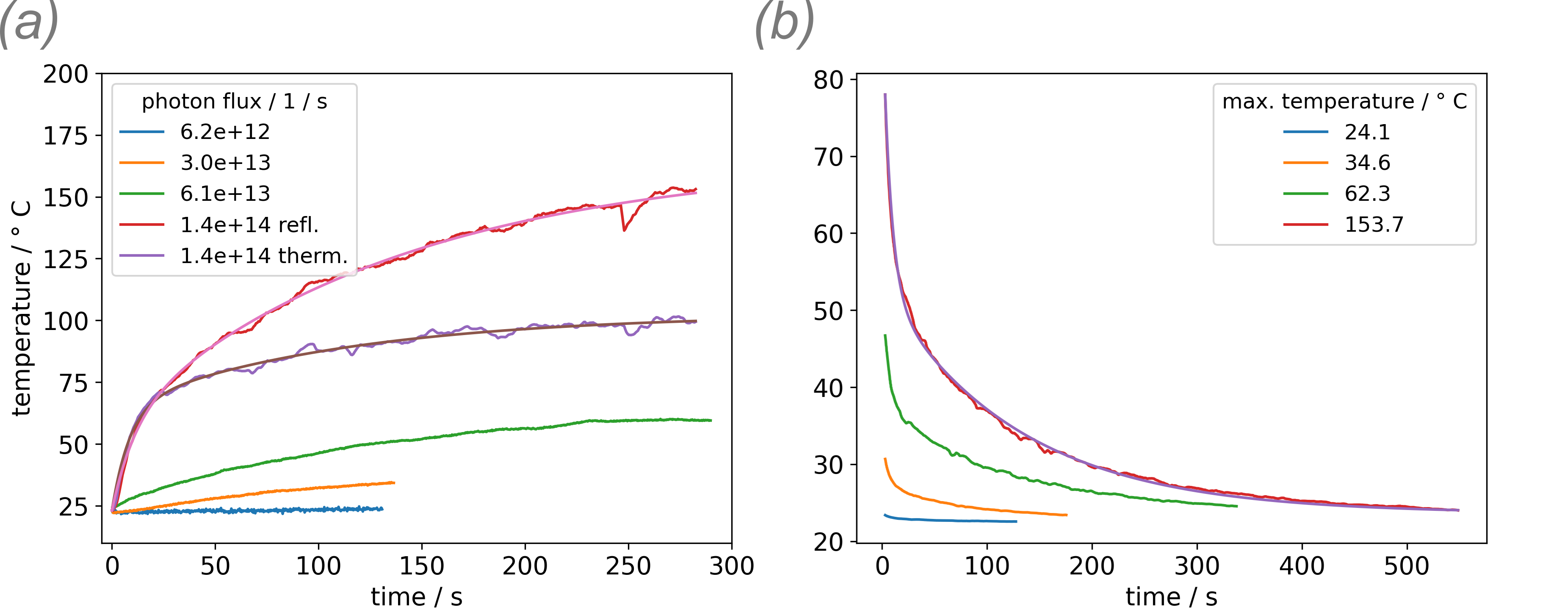}
    \caption{Temperature evolution during beam heating (a) and  subsequent cooling (b) for an Aluminum sample measured with a 17 keV pink beam. The red heating curve and all cooling curves are measured by a thermocouple while the other heating curves are measured by determining the shifts in the diffraction angle. The curves related to the highest photon flux are fitted to double exponential fit functions.}
    \label{beamheating}
    \end{center}
\end{figure}

To better understand such effects, we have focused on a mm-sized metal sample. For this, the characteristic internal diffusion time may be of order 1 ms, while the external heat transfer - e.g. to air - may be on the order of seconds or minutes. In the work of \citeasnoun{Bright2021}, heating and passive cooling at the ID11 beamline of the ESRF were described in terms of a lumped thermodynamic model. We employed a similar approach to DFXM data at the ID03 beamline. Specifically, we investigated the effect of pink beam exposure on the temperature of a 1.1 mm x 0.5 mm x 0.5 mm sized Aluminum sample. The sample was illuminated with a box beam characterized by an area of approximately 0.19 $\mathrm{mm^2}$ at 17 keV. The incident beam had a photon flux of $1.615 \times 10^{14}$ photons/s (assuming $8.5 \times 10^{14}$ photons/s for 1 mm x 1 mm slits). The cooling process was completely passive; no artificial cooling of the sample was employed. For this reason, sample cooling was due only to the heat exchange with the surrounding air at room temperature, the heat transfer inside the sample, and thermal radiation. The sample was mounted on a metallic sample mount via ceramic glue. The interface between the sample and the mount was limited, but its contribution to the cooling process is most probably not negligible. During heating and cooling processes, the temperature was measured with a thermocouple attached to the sample in a region not exposed to the beam. The temperature of the illuminated subvolume was also calculated by tracking the $2\theta$ angle of the 111 reflection of one grain in the far-field detector. This permitted an accurate measurement of the heating temperature. Measurements were repeated by varying the exposure time between 0.01 s, 0.05 s, 0.1 s and 0.5 s. The beamshutter was closed in between frame acquisitions for a certain amount of time, so the effective photon flux varied with the exposure time. The mean shutter closing times for the different exposure times were 0.25 s, 0.22 s, 0.17 s and 0.07 s, respectively. The effective photon flux was calculated by multiplying the photon flux on the sample of $1.615 \times 10^{14}$ photons/s by the ratio of the exposure time and shutter closing times. 
Results are shown in Fig.~\ref{beamheating}. The discrepancy during heating between the time evolution of the thermometer (purple curve) and the probe of the average temperature in the illuminated part of the sample (red curve) indicates that several processes with different time constants were at play, such as heterogeneous heat transfer inside the sample and transfer to the surrounding air. The curves can be well approximated by a double exponential function of the form:
\[
T(t) = T_{\mathrm{max},1} \cdot \left(1 - \exp\left(-\frac{t}{\tau_1}\right)\right)
      + T_{\mathrm{max},2} \cdot \left(1 - \exp\left(-\frac{t}{\tau_2}\right)\right),
\]
 for both heating and cooling. The lumped model predicts exponential temperature rise and decay over a single timescale. For the curves in Fig.~\ref{beamheating}, this model breaks down and two time-scales must be introduced to fit the data. 
 Moreover, at around 250 s a local minimum in temperature can be seen for the highest flux curve. Here, the time between two frames was about 2 s instead of the usual 0.5 s. This leads to a cooling of about 10 \textdegree C. The same result is apparent from the initial cooling in
Fig~\ref{beamheating} (b).

Since only a part of the sample was illuminated (unlike in \citeasnoun{Bright2021}) and the heat was transferred inside the sample, several different cooling mechanisms with different timescales contributed. As a result, the lumped model approach is not optimal for this system.

These results showcase three challenges for pink beam operation in metals:
\begin{itemize}
    \item The steady-state temperature in the unattenuated pink beam at 17 keV - using prefocusing by means of a transfocator but no condenser -  is several hundred degrees and may lead to sample melting. 
    \item The time to reach equilibrium may be 5-10 minutes (the red curve in Fig~\ref{beamheating} (a) has not yet saturated).
    \item The cooling time is sufficiently rapid (in the order of $\sim$ 10  \textdegree C / s ) that for example extended motor movements between single scans (with the shutter closed) can lead to a significant temperature drop and potential changes in microstructure and stress state.
\end{itemize}

\section {Discussion}
The results of this study show that the suitability of pDFXM depends on how the trade-offs between time, spatial, and angular resolution impact specific materials science applications. 

In particular, our findings on beam damage suggest that a protocol for heat distribution and \emph{in situ} temperature characterization may be useful for each specific scientific case. As we demonstrated, monitoring the sample temperature via a thermocouple or an infrared camera for contactless measurements, as well as registering the movement of a reflection on a test scan, can be practical solutions. 

We foresee that for many science cases it is of relevance to swap between the use of a pink beam (overview scans over large orientation ranges and/or large sample volumes, fast acquisition during time series) and the use of a monochromatic beam (high spatial and angular resolution in general, strain scanning). At the ID03 beamline, switching between the two illumination schemes only takes a few minutes \cite{yildirim2025pink}. 

An additional advantage of pink beam operation may be to alleviate problems with dynamical diffraction - similar to work in EBSD \cite {Winkelmann2010} and pulsed electron microscopy. This is potentially an important argument for the use of pDFXM, but is outside the scope of this work. 

Regarding chromatic aberration, we find dark field microscopy to be very different from bright field microscopy. The fact that the sample effectively acts as a monochromator implies that for a perfect crystal, which will never be the case in reality, and for a near parallel incident beam there is no aberration. This  is of interest for magnified topography studies, e.g. inspection of rare  defects in large single crystals.  The introduction of a condenser, which is indispensable for many DFXM studies, leads to an anisotropic broadening, which still can be modeled for perfect crystals.  For imperfect crystals, in particular plastically deformed ones, it will however be difficult to include the effects of aberration in forward modeling as it is not only strongly sample-dependent but also dependent on the degree of mosaic spread within the individual voxel. For operation we propose to compare pDFXM and normal DFXM images directly to learn if the aberration is of concern.

In the conventional monochromatic DFXM, the $\phi$ motor - or alternatively a base tilt  $\mu$ - is used to map dislocations because of the high resolution in the rocking direction that provides good weak-beam conditions. In the case of the pink line beam illumination, one needs to increase the range of the rocking scan to reach the same intensity contrast as the monochromatic case. Nevertheless, as demonstrated in Fig.~\ref{fig:weakbeam}, the use of a pink beam ensures enough contrast in the weak-beam region for a clear detection of dislocations. 

Throughout this paper, we have assumed the use of CRL based optics. Multilayer Laue Lenses, MLLs, are an interesting alternative, due to the significantly larger NA, up to 0.016 for 20 keV operation \cite{Braun2013, Bajt2017, Chapman2021}. At the same time, manufacturing errors can be reduced, implying that spatial resolution can be greatly improved in relation to CRL operation. For classical tomography, in bright field, focal spots as small as 2.9 $\times $ 2.8 nm$^2$ have been demonstrated at a photon energy of 17.5 keV \cite{Dresselhaus2024}. Two crosslinked MLLs may be used as an objective for DFXM \cite{Murray2019,Kutsal2019}. The specifics of both monochromatic and pink beam properties of a DFXM instrument with such an objective is highly interesting but has yet to be derived theoretically. However, a basic understanding can be obtained from the equations derived here from the CRL case, simply by inserting the larger NA of the MLL. As an example, for the reciprocal space resolution function the $\sigma_a$ terms become dominant in the monochromatic beam version of Eqs. \ref{eq-q_rock}, \ref{eq-q_roll} an \ref{eq-q_par} while the pink beam will be characterized by the fact that nearly the entire energy spread can be transmitted through the lens.

\section {Conclusion}


This work provides rigorous evaluation of the benefits and limitations associated with the use of pDFXM. We have extended the geometric optics framework to account for chromatic effects and applied the numerical formalism presented in \cite{Borgi2024} for forward modeling under polychromatic conditions.

The degradation of reciprocal-space resolution in the rocking and longitudinal directions is a generic feature, independent of sample type. This limits the angular sensitivity for high-resolution strain mapping but can, conversely, facilitate more efficient intensity integration over broad reflection ranges. The corresponding effects in direct space are more complex: the use of a condenser introduces an asymmetric chromatic aberration that is approximately homogeneous across the field of view, while local deformation amplifies the blur in a sample-dependent manner. Consequently, forward modeling of pDFXM images becomes increasingly difficult for plastically deformed crystals, where the mosaic spread varies strongly across the illuminated volume.

Despite these limitations, the weak-beam conditions achievable with pDFXM are particularly attractive for studying heavily deformed grains. Such grains exhibit large orientation and strain spreads but often produce weak diffraction signals due to their high mosaicity and degraded crystal quality. In these cases, pink beam provides a unique opportunity to visualize and map bulk deformation structures with a still superior angular resolution compared with electron backscatter diffraction (EBSD).

Beam heating remains a practical concern. Given that typical DFXM scans involve repetitive illumination cycles, thermal load management must be considered carefully for each sample. Increasing the photon energy will alleviate these effects, and the ongoing transition from beryllium to diamond compound refractive lenses (CRLs) will support this development by enabling higher-energy operation with reduced absorption.

The observed decrease in spatial resolution can nevertheless be well described in the presence of dislocation structures. 

Finally, the combination of pink beam with magnified topotomography holds great promise for time-resolved three-dimensional imaging \cite{shukla2025}. The broad energy bandwidth associated with the pink beam enables high flux and fast acquisition, ideal for dynamic experiments. Fully exploiting this approach will be a grand challenge for future studies, demanding further development in both experimental design and analysis methods.

\section{Acknowledgment}
The authors thank Sina Borgi for help with the reciprocal space resolution code and ESRF for providing the beamtime at ID03.  MLB and HFP acknowledge financial support from the ERC Advanced Grant nr 885022 and from the Danish ESS lighthouse on hard materials in 3D, SOLID. CY acknowledges support from the ERC Starting Grant nr 10116911. 
C.Y. acknowledges the technical help provided by
H. Isern and T. Dufrane during the experiments.

\referencelist[references]

\end{document}